# Direct imaging of quantum interference and Non-Abelian entanglement in Hopfion: an magnetic soliton possess loop-like anyonic properties

**Jiawei Dong[1,5], Xin Zhang[1], Hailong Shen[1], Haoyu Wu[1], Zhenyu Ma[2], Yong Deng[3], Wenyu Hu[4], Wenbin Qiu[7], Bo Liu[2], Xiaoyi Wang[1#], Yihan Wang[2#], Longqing Chen[2#], Yang Qiu[5#], Jian Ma[1], Xudong Cui[6], Kun Zhang[2] and Pierre Ruterana[8].**

1 College of Electronie&Infermation, Southwest Minzu University, State Ethnic Affairs Commission, Chengdu 610041, China.

2 Key Laboratory of Radiation Physics and Technology of the Ministry of Education, Institute of Nuclear Science and Technology, Sichuan University, Chengdu 610064, China.

3 School of Materials Science and Engineering, Key Laboratory for Polymeric Composite and Functional Materials of Ministry of Education, Institute of Green Chemistry and Molecular Engineering, Sun Yat-sen University, Guangzhou 510275 P.R. China; School of Physical Sciences, Great Bay University, Dongguan, 523000, P. R.China.

4 College of Science, Southern University of Science and Technology, Shenzhen 518056, China.

5 Pico center, SUSTech Core Research Facilities, Southern University of Science and Technology, Shenzhen 518055, China.

6 The School of Optoelectronic Science and Engineering, University of Electronic Science and Technology of China, Chengdu, 611731, China.

7 Department of Fundamental Courses, Wuxi Institute of Technology, WuXi 214121, China.

8 CIMAP, UMR 6252, CNRS-ENSICAEN-CEA-UCBN, 14050 Caen, cedex, France.

Correspondence： Xiaoyi Wang 80300024@swun.edu.cn,
Yang Qiu qiuy@sustech.edu.cn,
Yihan Wang yhwang027@scu.edu.cn,
Longqing Chen chenlongqing@scu.edu.cn.

## Abstract

As topologically protected three-dimensional solitons, magnetic hopfions have garnered considerable interest for spintronic applications due to their particle-like stability and localized magnetization topology. Theoretical frameworks suggest that tailored perpendicular magnetic anisotropy in frustrated/chiral magnets could stabilize skyrmionic states, with twisted skyrmion strings forming closed-loop hopfion structures through geometric confinement. While real-space imaging of hopfions has recently been achieved [4], the quantum-mechanical consequences of their braiding dynamics—particularly quantum interference and entanglement between hopfion rings and embedded skyrmion strings—remain experimentally unverified, hindering progress in understanding their quantum topological phenomena.

Here, we report the critical evidence arises from rotational-axis-asymmetric quantum interferograms of hopfion. i) marked by symmetry reduction from $S(\infty)$ to C4 (in-axis) and Cs to C1 (off-axis) under field modulation (200 – -200mT)—which contrast fundamentally with Aharonov-Bohm-type interference in skyrmion lattices. ii) Field-tuned tomographic imaging reveals topological charge fractionalization ($Q=1 \rightarrow Q=1/2$) at external magnetic strength $H$ = -50 mT, manifesting as vortex-to-crescent soliton transitions that obey SU(2)-mediated meronic edge state dynamics. iii) Coherent skyrmion-string deformations—distinct from independent oscillations in multi-skyrmion systems—demonstrate entanglement-driven soliton correlations.

This work provides the first experimental elucidation of quantum topological effects in individual hopfions, establishing their potential as building blocks for three-dimensional topological quantum spintronics. The observed Non-Abelian characteristics suggest pathways toward fault-tolerant quantum operations through controlled hopfion braiding in engineered magnetic metamaterials.

**Introduction**

Three-dimensional magnetic hopfions, topologically protected solitons characterized by closed-loop skyrmionic textures, have emerged as critical candidates for beyond-von Neumann computing architectures [1]. Their stability stems from the interplay between localized magnetization confinement and global Hopf invariant

conservation—a combination enabling robust three-dimensional (3D) information encoding essential for high-density spintronic implementations.

First predicted in 1975 through Faddeev's nonlinear σ-model describing twisted skyrmion-string closures [2], hopfions remained experimentally elusive for decades due to the challenge in access of spatial resolution down to nanometer scale. The field witnessed transformative progress in 2021 when soft X-ray correlated microscopy achieved the first real-space detection of stabilized hopfions in synthetic Pt/Co/Ir multilayers [3], followed by Lorentz microscopy imaging of hopfion in B20-type FeGe films in 2023—establishing their existence across diverse chiral magnet systems [4]. Despite these milestones, fundamental questions persist regarding hopfion quantum dynamics. Conventional Special Orthogonal group in 3 dimension (SO(3)) [5] restrict the analysis of integer-spin quasiparticles (e.g., Q=±1 skyrmions), while the critical external magnetic excitation ($H_c$) for fractional-spin states (e.g., Q=±1/2 meron) in Special Unitary 2×2 matrix group (SU(2)) remains unexplored in hopfion. Particularly, braiding operations between hopfion rings and constituent skyrmion strings should generate path-ordering dependent Non-Abelian Berry phases (Fig. 2b), theoretically enabling:

1）Symmetry destruction of Quantum interference driven by rotation operated Non-Abelian phase matrix

2）SU(2)-symmetric quantum entanglement through off-diagonal phase matrix elements

3）Half-quantized solitonic excitations yielding loop-anyonic statistics via symmetry breaking of quantum interference

Current models fail to account for these quantum-coherent phenomena, creating a critical knowledge gap between hopfion magnetization and their potential as topologically protected qubits.

Here we demonstrate Non-Abelian quantum effects in hopfion through Lorentz transmission electron microscopy (LTEM) operating in Fresnel-defocus imaging mode. Stabilized in a disordered FeCrNiMn high-entropy alloy with surface oxidization, the hopfion arises from interfacial engineering of perpendicular magnetic anisotropy (PMA) and Dzyaloshinskii-Moriya interaction (DMI) gradients [3]. Depth-resolved magnetization mapping, reconstructed through the transport of intensity equation (TIE) formalism [6], revealing the topological charge equality between constituent skyrmion strings ($Q=-1$) and hopfion ring ($Q=-1$) as well as the $\mathbb{R}^3$-space braiding signatures of observed hopfions.

Notably, rotational-axis-asymmetric quantum interferogram emerge in hopfions, contrasting with the symmetrical Aharonov-Bohm-type interference in multi-skyrmion systems. Furthermore, field-tunable interferometry shows symmetry reduction from $S(\infty)$ to C4 under external magetic excitation from 200 mT→-200 mT, directly evidencing Non-Abelian Berry phase accumulation—a phenomenon absent in conventional skyrmion bundles [7]. The evidences of quantum entanglement manifest as:

1) Field-synchronized soliton dynamics: Coherent skyrmion-string deformations in hopfions versus independent size oscillations in multi-skyrmion lattices

2) Topological charge fractionalization: Field-induced $Q=1→Q=1/2$ transitions (vortex→crescent morphologies) at $\boldsymbol{H_c}$=-50 mT, aligning with SU(2) group predictions for meronic edge states

While current braiding theories in 2D systems inadequately describe the loop anyonic statistics of hopfion, the data demand development of higher-order braid group statistics to capture the SU(2)-mediated anyonic phase transitions and Non-Abelian holonomy in 3D soliton networks. This work establishes the disordering of high-entropy alloys could potentially been a platforms for engineering topologically protected quantum qubits, with implications spanning skyrmion braids [8], solitonium lattices [9], providing a novel guideline for architecturing the neuromorphic spintronic [10].

**Experiment**

The $TiO_2$-dispersion-strengthened FeCrNiMn high-entropy alloy was fabricated through gas-atomized synthesis (Ar atmosphere, 99.999% purity) followed by the

Mechanical alloying and Consolidation, the details of material synthesis can be found in our recent publication [11]. 6 MeV Au⁺ ion irradiation (fluence 5×10¹⁵ ions/cm², 350°C) was performed using a tandem accelerator in the Key Laboratory of Radiation Physics and Technology of Sichuan University, generating ballistic dissolution [12] induced disorder critical for topological defect stabilization. the TEM specimen was prepared by focus ion beam (FIB) thinning using Thermo-Fisher Helios 600i, a 30 kV Ga+ ion beam with beam current of 0.23 nA-9.3 nA was used to extract FIB lamella, the final thinning of the lamella was performed with a 2 kV Ga⁺ ion beam, where a 15 pA beam current in order to reduce the beam damage when going down to ~100 nm. A subsequent surface polishing was conducted in a Fischione Nanomill 1040, where the beam current and the accelerating voltage were set as 150pA and 900 eV to to eliminate the residual surface damage from FIB fabrication. Subsequently, the TEM specimen was transported into a vacuum chamber with Atmospheric Pressure of ~10⁻¹ Torr, oxidation proceeds selectively in the disordered regions, leaving the crystalline domains intact. An In-situ ±200 mT excitation (50 mT steps) synchronized with Fresnel-mode LTEM (Talos 200X, 200 kV) was used to probe the micromagnetic field with ±70° angular sampling using Fischione 360° tomographic holder for 3D spin reconstruction. The Pylorentz Transport-of-intensity equation (TIE) formalism [7] reconstructing 3D spin textures through discrete Fourier analysis of defocus series (±10 μm), drift correction and noise suppression was included in the software.

**Theory**

The braiding dynamics of skyrmion strings in three-dimensional demonstrates Non-Abelian holonomy when forming Hopfion topology. The emergent noncommutative geometry modifies the fundamental commutator algebra [13,14]:

$$[X^\mu, P^\nu] = i\hbar\delta^{\mu\nu} + ie^{\mu\nu p}H_p \quad (\mu, \nu, p \epsilon \{x, y, z\}) \quad (1)$$

Where $H_p = Q_H \frac{\partial \phi}{\partial p}$ encodes the hopfion charge density and $\phi$ is the azimuthal twist angle. The SU(2) gauge field enables the emergence of fractionalized topological charges $Q_H$ ( $Q_H = \frac{1}{4\pi} \int_{S^3} H ds \ \epsilon \ Z$ ), a hallmark of Non-Abelian Hopfion configurations, the interplay of Berry flux quantization and manifold topology underpins the breakdown of integer quantization constraints by SO(3) gauge group. The $\nabla \cdot H = 4\pi Q_H$ enforces topological charge conservation. Considering the Non-Abelian Berry connection manifests as an SU(2) gauge field [15,16]:

$$A = \frac{Q_H}{2r^2}(\sigma_y^x - \sigma_x^y)d\phi + \frac{\chi}{r^3}\sigma_Z Z dr \tag{2}$$

Where $r = \sqrt{x^2 + y^2 + z^2}$, $\sigma_i$ are Pauli matrices, $\chi$ denotes skyrmion helicity. Under rotational detection, the holonomy operator decomposes as [16,17]:

$$\phi_{N_A} = p^{3D} exp\left[i\phi(MAM^\dagger)_\mu dx^\mu\right] + \frac{Q_H}{2}\int_S \Omega_{ij}dx^i \wedge dx^j \tag{3}$$

Where M is the rotation matrix describes sample tilt orientation and $\Omega ij$ is the curvature two-form, $\Delta\Omega_{ij} = \partial_i A_j - \partial_j A_i + i[A_i, A_j]$, $p^{3D}$ denotes the extended Yang-Mills path-ordering operator in 3 dimension [18]. The interference intensity can be then written as:

$$I \propto 1 + Re[Tr(\Delta\phi_{N_A})] + (-1)^{n_{twist}}cos(arg\Delta\phi_{Abelian}) \tag{4}$$

Where Tr($\Delta\phi_{N_A}$) denote the matrix trace operation of Non-Abelian phase matrix and $n_{twist}$ represents the number of topological braiding operations. The interference intensity reveals symmetry transitions by Non-Abelian phase accumulation. The key insight quantum phenomena includes the quantum entanglement and anyonic statistics, i) for $Q_H \neq 0$, the Non-Abelian phase matrix manifest a non-diagonal element, preventing hopfion wavefunction factorization due to the violation of Bell inequality. Otherwise, The Abelian limit ($Q_H=0$) suppresses quantum entanglement via the degeneracy of Non-Abelian phase difference, eg. $\lim_{Q_H \to 0} \Delta\phi_{N_A} \to dig(e^{i\phi_1}, e^{i\phi_2}, \cdots, e^{i\phi_n})$.

ii) The homotopy charge $Q_H$ governs the emergent excitations in SU(2) group space.

$$\Psi_{edge} \sim \begin{matrix} Ising\ anyon & Q_H \epsilon 2Z + 1 \\ Fibonaci\ anyon & Q_H \epsilon 2Z \end{matrix} \tag{5}$$

Current 2D meron theories[19] fail to capture 3D statistics, necessitating a comprehensive higher braid group representations.

**Results and discussion**

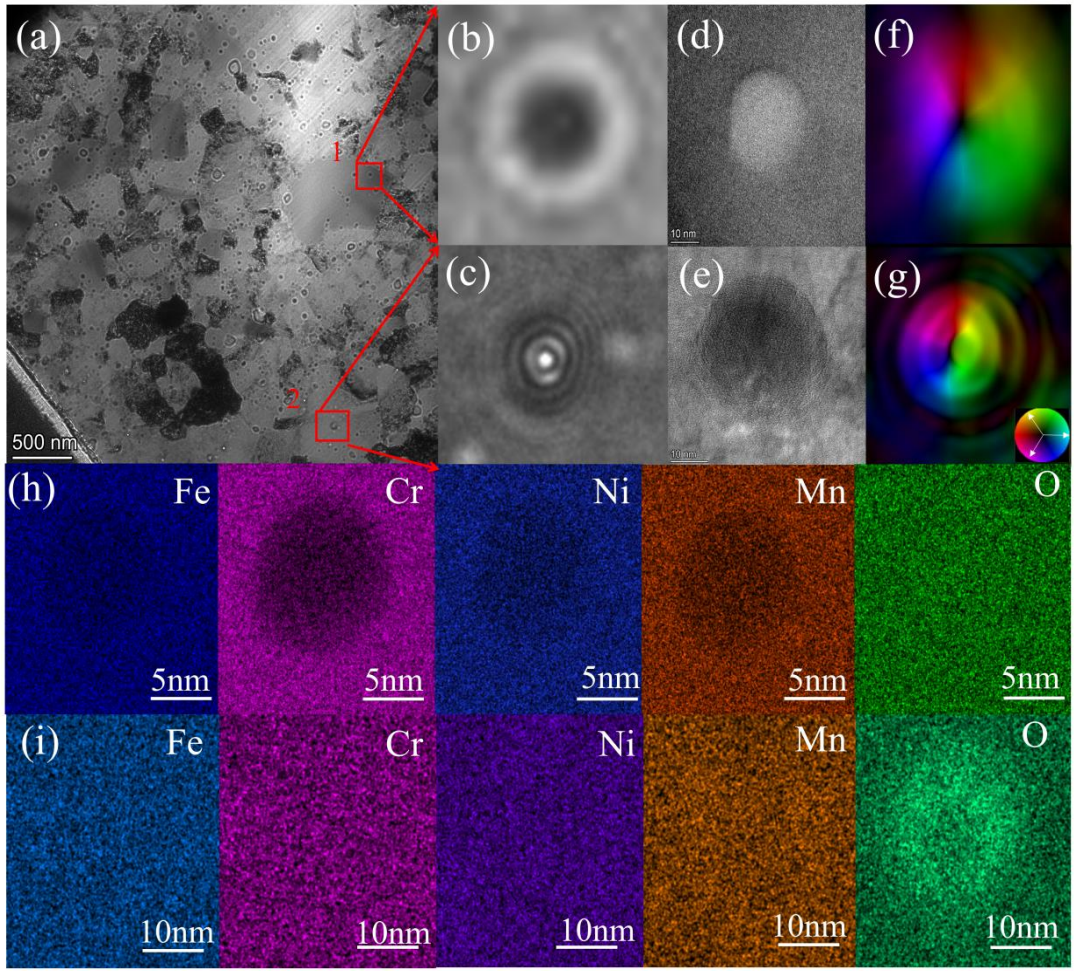

Fig.1 (a) Lorentz electron microscope image; (b) A single skyrmion Lorenz image; (c) Hopfion Lorenz image ; (d) High-resolution electron microscope images of individual skyrmion;(e) High-resolution electron microscope images of hopfion; (f) Magnetic domains of central sections of skyrmion; (g) Magnetic domains of central sections of hopfion;   elemental maps of region of skyrmion(h) and Hopfion (i)

The existence of hopfions must satisfy strict symmetry-breaking conditions than of skyrmions. The skyrmion phase can be stabilized in disordered systems with vanishing magnetic anisotropy, where Dzyaloshinskii-Moriya interaction (DMI) dominates over Heisenberg exchange [20]. while the Hopfion phase requires concurrent interface-enhanced PMA and bulk DMI to generate helical confinement [3]. The radiation-induced disorder in FeCrNiMn high-entropy alloy creates a magnetic frustration landscape, in non-oxidized regions, the preservation of isotropic exchange supports 2D skyrmions formation, while for the surface oxidized region, the develop PMA gradients at interface through oxygen-mediated orbital hybridization, enabling 3D spin modulation to produce hopfion. To resolve oxidation-modulated topological phase transition in disordered region, the atomic images with their corresponding elemental maps were shown in Fig 1. In Fig. 1d-e, both ion-irradiated region exhibits randomly distributed lattice fringes with a closely resembling strain distribution (Fig. S1). However, the elemental maps reveal that only region 2 was properly oxidized. In Figure S1, the electron energy loss spectra (EELS) analysis revealed a distinct oxygen K-edge signature exclusively localized within the hopfion regions, with spatial correlation confirmed by elemental mapping in Fig. 1h-i.

Driven by -50 mT external magnetic excitation, the TIE-reconstructed spin texture reveals bloch-type skyrmion in prestine disordered region[21], while in the oxidized region, the Toroidal confinement of Bloch skyrmion was observed (Fig.1 f), where the conserved winding number coherence [4] exhibits a magnetization configuration in region 2 distinct from skyrmion bags [22]. To verify the hopfion protocol, three-dimensional spin reconstruction was used to confirm closed loops in $\mathbb{R}^3$ space. To unambiguously establish that hopfion formation in FeCrMnNi alloys is exclusively mediated by surface oxidation, the comparative studies of skyrmion and hopfion stabilization in ion-damaged crystalline Cr is provided (Fig. S1). The strain field analysis (Fig. S1) reveals nearly identical strain distributions in both skyrmion- and hopfion-stabilizing regions. Crucially, electron energy-loss spectroscopy (EELS) detects oxygen K-edge signals solely within hopfion domains (Fig. S1g). These findings demonstrate remarkable consistency with the skyrmion- and hopfion-stabilization mechanisms observed in FeCrMnNi systems, providing evidence for oxidation-driven topological phase transitions in disordered materials.

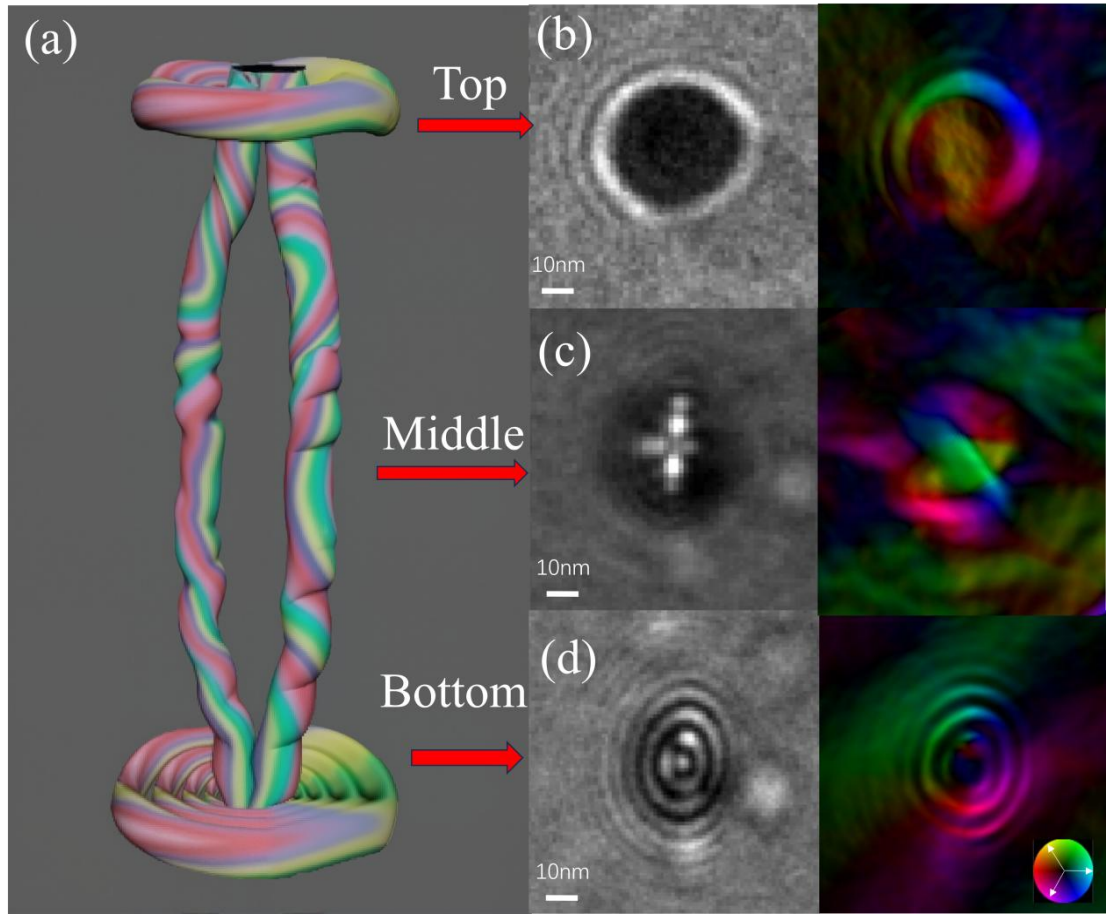

Fig.2 (a) Schematic diagram of Hopfion 3D reconstruction structure; (b) Top slice lorentzian microscope and magnetic domain mapping of Hopfion; (c) Lorentzian electron microscope images and magnetic domains of central sections of Hopfion; (d) Bottom slice lorentzian microscope and magnetic domain mapping of Hopfion.

To resolve the hopfion's topological microstructure, we employed defocus-series tomography (the complete tomographic LTEM images are shown in Fig S2 via quantitative Lorentz transmission electron microscopy (LTEM) , a $\Delta f = \pm 10$ μm step size was used to perform depth-resolved spin texture reconstruction by TIE. In Fig 2 b-d, identical structure has been revealed at each depths of the magnetic domain, where the spin texture demonstrates a conserved topological charge of vortices at surface layer and bulk region. In intermediate regime in Fig 2c, an orthogonally intersecting flux-tube structures was noticed, where the magnetization configuration exhibits two separated magnetic vortex. As these topological defect fingerprints are absent from skyrmion bundle tomography (Fig S3), the 3D magnetization field evidences a depth related berry phase accumulation, confirming the Non-Abelian braiding statistics of hopfion in $\mathbb{R}^3$ space.

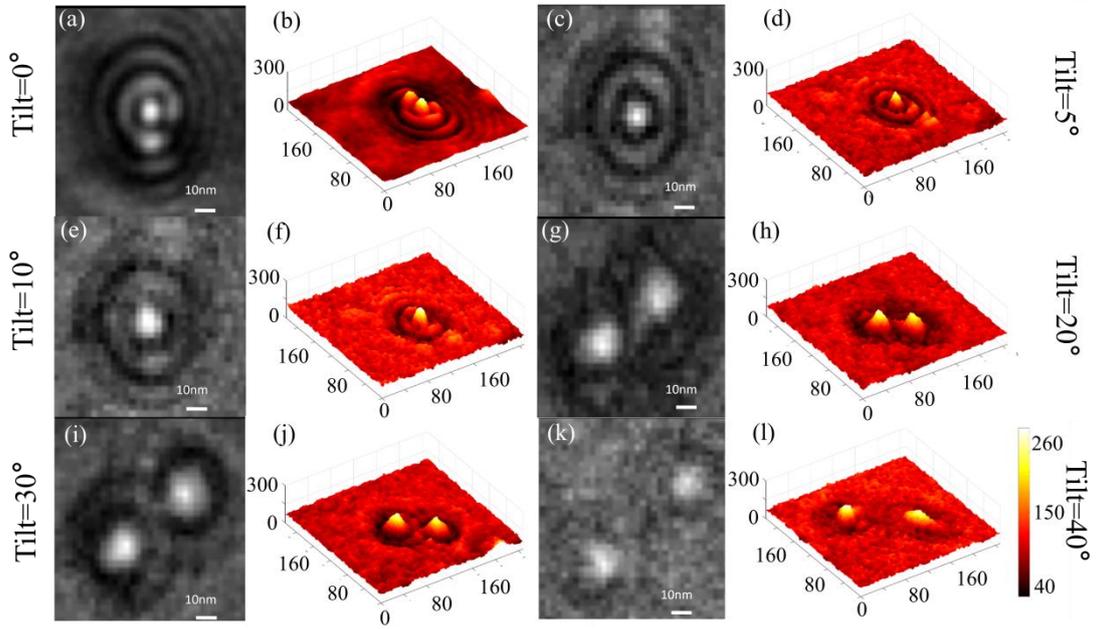

Fig 3 (a),(c),(e),(g),(i) and (k): the Lorentz TEM imags and reconstructed magnetic domain maps of the hopfion configurations under different tilt angles (0°, 5°, 10°, 20°, 30° and 40°);; (b),(d),(f),(h),(j) and (l): 3D reconstruction of Hopfions from different angles (0°, 5°, 10°, 20°, 30° and 40°)

Manifested by equation 3, the hopfion exhibits rotation-dependent decoherence of quantum interference at certain path-ordered contribution, which fundamentally distinct from multi-skyrmion systems. In Fig 3a-k, by minimizing the path-ordering endowment at zero external field (unperturbation of skyrmion strings braiding at 0 mT), tilt-induced symmetry reduction follows interferogram topological transition, allowing the S(∞)→ Cs →C1 (circle→semicircle→arc shaped interference fringes) symmetry breaking been observed in hopfion, which evidences the rotation induced Non-Abelian phase accumulation in equation 3. While in Fig S4, the tilt-immunity of Abelian phases of skyrmion pair guide the symmetry protection of interferogram. To rigorously exclude thermal decoherence effects [23] in hopfion interferometry, the tomographic LTEM imaging was performed under cryogenic conditions (~90 K), achieving the thermal drift suppression and phonon freezing. The dual-temperature protocol suggests a symmetry presistence of interference pattern and conserved interference visibility $V = \frac{I_{max}-I_{min}}{I_{max}+I_{min}}$ (Fig S5) during the temperature drop, referring a negligible thermal influence on hopfion interferogram. Following rigorous exclusion of thermal decoherence, symmetry analysis reveals the Non-Abelian holonomy of hopfion interference patterns. The rotation-angle-dependent unitary matrix contributes the Non-Abelian phase accumulation in hopfion, while the interference of skyrmion pair was strictly protected by SO(3) gauge symmetry at *H*=0.

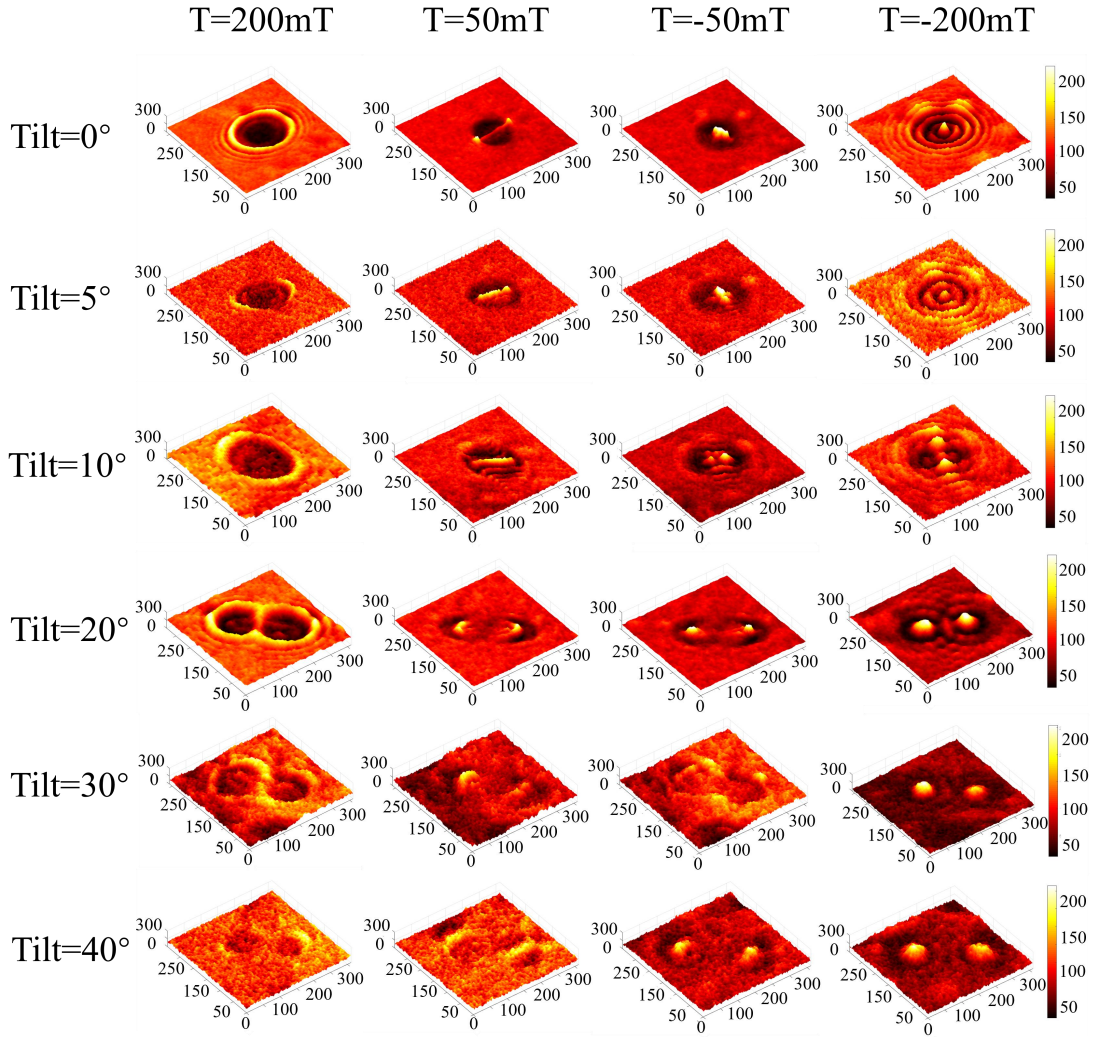

Fig 4: Three-dimensional interferogram of Hopfion driven by different field strengths.

Notably, equation 4 establishes that the rotation-angle-dependent unitary matrix and path-ordering contribution governs path-ordered holonomy, the dissection experiments was designed to address the contribution of path-ordering operator ($p^{3D}$). The interplay between magnetic excitation and topological braiding geometry was systematically investigated via four-quadrant magnetic excitation cycle, where the 0

mT→200 mT→0 mT→-200 mT→0 mT Magnetic field trajectory was used to probe the hopfion braiding dynamics (Fig 4). In Fig 4, field-driven interference symmetry transition was resolved, particularly for field sweep from 200 mT to -200 mT, an explicit symmetry hierarchy transition was observed. The In-axis interferogram response demonstrates a symmetric reduction (from S(∞) to C4), while the off-axis geometric perturbation scheme reveals a symmetry transition from Cs (semicircle) to C1 (crescent shape). As systematically compared with skyrmion bundle in Fig S4 (absent of symmetry breaking), the hopfion symmetry alteration emerges from braiding-governed quantum geometry.

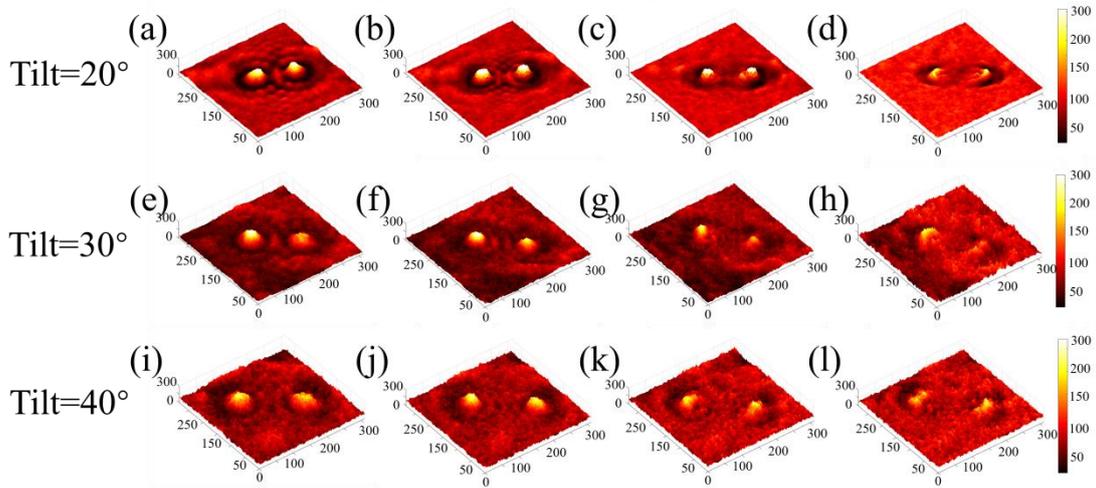

Fig 5 (a)-(l) A pair of skyrmion strings synchronize the vortex to crescent transition at 20º 30º and 40º.

The skyrmion strings entanglement nexus is encoded in the off-diagonal terms of the Non-Abelian phase matrix, as resolved through field interferometric tomography (Fig. 4). The field-programmed braiding dynamics is able to guide the magnetic texture synchronization, particularly at high angle (20º-40º) tomographic imaging in Fig 5, where the field-locking of a pair of skyrmion strings perform a synchronous vortex-to-crescent transition. However, through multi-skyrmion entanglement null test in Fig S6, only the size-dependent decoherence was noticed between each skyrmions. These observation prove the hopfion-mediate quantum entanglement originates from Non-Abelian braiding topology. Of note, a field-induced vortex-crescent metamorphosis emerges as a signature of dimensional hierarchy in quantum topology, captured through critical-field magnetic tomography in Fig 5, The observed soliton-meron transition induces a fractionalized charge regime, where the halved topological charge satisfies the anyonic fusion rule. However, as the anyonic signature was restricted in 2D system, a compelete loop braiding theory is required to explain our observation, eg,. loop-like anyons.

**Conclusion**

In this study, we have unearthed Non-Abelian quantum effects within hopfions, leveraging the precision of Lorentz transmission electron microscopy (LTEM) in Fresnel-defocus imaging mode. Our investigation, confirm the stabilization of hopfion in a disordered FeCrNiMn high-entropy alloy with surface oxidization. Depth-resolved magnetization mapping, reconstructed via the transport of intensity equation (TIE) formalism, reveals intriguing topological insights of topological charge equivalence and $\mathbb{R}^3$-Space braiding signatures of target hopfion. Notably, hopfions exhibit rotational-axis-asymmetric quantum interferograms, contrasting sharply with the symmetrical Aharonov-Bohm-type interference observed in multi-skyrmion systems. Field-tunable tomographic interferometry further demonstrates a symmetry reduction from $S(\infty)$ to C4 (in-axis) and Cs to C1 (off-axis) under external magnetic excitation, directly evidencing the accumulation of Non-Abelian Berry phases—a phenomenon absent in conventional skyrmion bundles. Manifestations of quantum entanglement includes the coherent skyrmion-strings deformation within hopfion, as opposed to independent size oscillations in multi-skyrmion lattices, underscore the entangled nature of these structures. Of note, the field-driven magnetic tomography suggest the existence of Topological Charge Fractionalization in hopfion, the field-induced transitions from Q=1 (vortex morphologies) to Q=1/2 (crescent morphologies) at a critical field H=-50 mT align with SU(2) group predictions for meronic edge states, indicating fractionalization of topological charge. While existing braiding theories inadequately describe the loop anyonic statistics observed in our study, our data necessitates the development of higher-order braid group statistics to capture SU(2)-mediated anyonic phase transitions and Non-Abelian holonomy in 3D soliton networks. This work establishes the disordering of high-entropy alloys as promising platforms for engineering topologically protected quantum qubits, providing a novel guideline for architecting neuromorphic spintronics. Our findings open avenues for future research in harnessing Non-Abelian quantum effects of magnetic hopfion for advanced quantum information processing and spintronic applications.

## Acknowledgments

This work was supported by the National Natural Science Foundation of China (NO. 52203232) and the Sichuan International Science and Technology Cooperation Program (Grant No. 2025YFHZ0049) and the institutional Joint Innovation Fund from Sichuan University and Nuclear Power Institute of China (SCU&NPIC-LHCX-25). Xiaoyi Wang gratefully acknowledges Prof. Xingwen Liu (College of Electrical Engineering, Southwest Minzu University) for insightful discussions on the mathematical modeling of hopfions.

**Supplemental material**

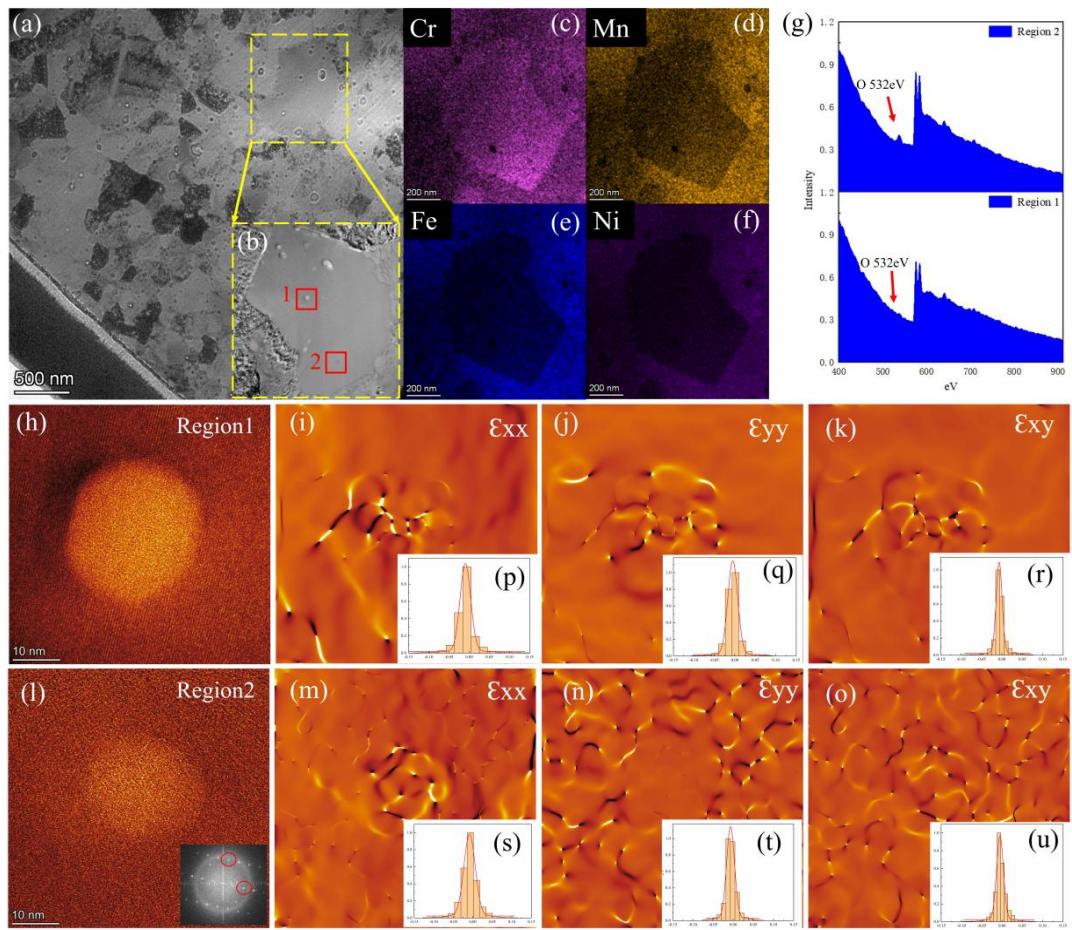

Fig.S1 (a) Lorentzian and (b) TEM image of skymion on crystalline Cr (region 1), and hopfion on crystalline Cr (region 2); (c)-(f) are the elemental maps of Cr; (g) is the EELS spectra of skyrmion and hopfion region in (a); The strain contours (h)-(k) and histogram statistics (p)-(r) near the hopfion position showcase the strain characteristics of a skyrmion in the same crystalline structure, with $\varepsilon_{xx}$ and $\varepsilon_{yy}$ representing principal strains in the (110) and (220) planes, respectively, and $\varepsilon_{xy}$ denoting the shear strain in the (220) plane.. The GPA lattice displacement maps (l)-(o) and corresponding strain statistics (s)-(u) illustrate the strain distributions of a hopfion on BCC crystalline Cr with same strain direction as skyrmion in (h)-(k).

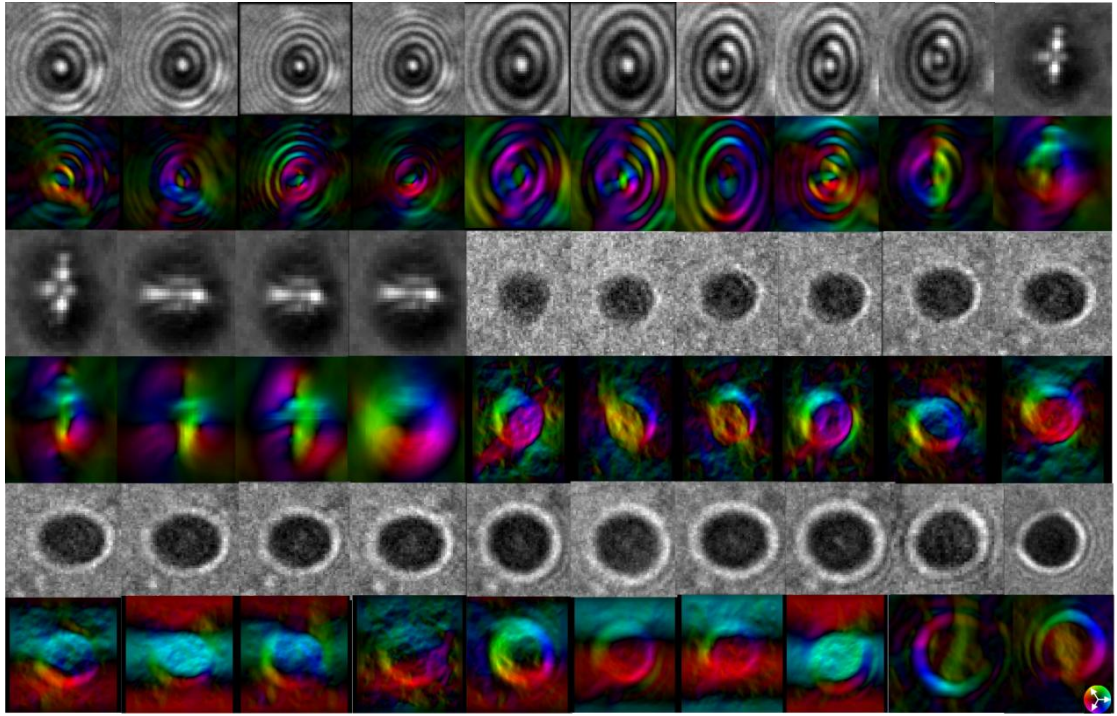

Fig.S2 Lorentzian electron microscope images and magnetic domain images of Hopfion at a step of defocus factor of 10μm.

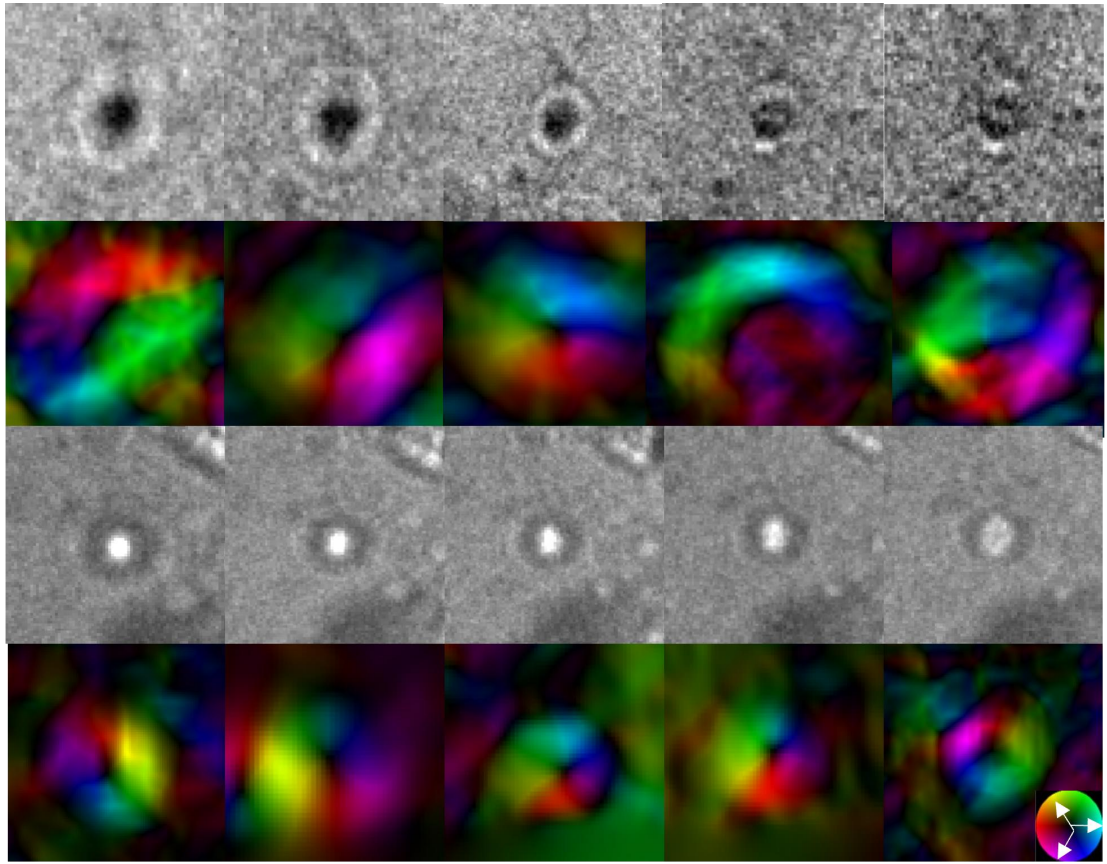

Fig.S3 Lorentzian electron microscope images and magnetic domain images of Skymion.

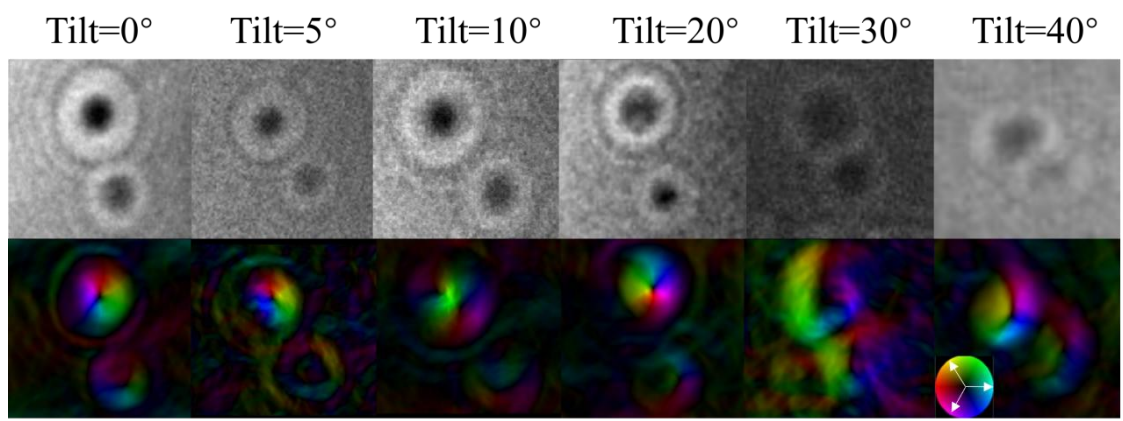

Fig.S4 Lorentzian electron microscope images and magnetic domains of Skymions at different tilt angles.

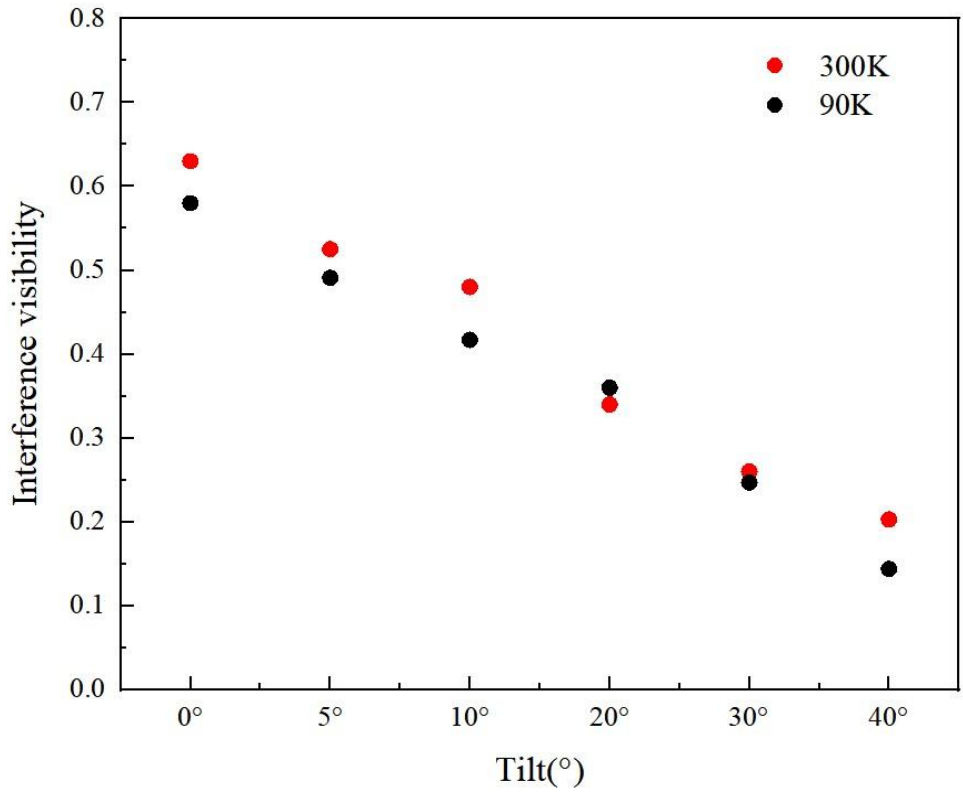

Fig.S5 Visibility of diffraction patterns in samples at 90K and 300K.

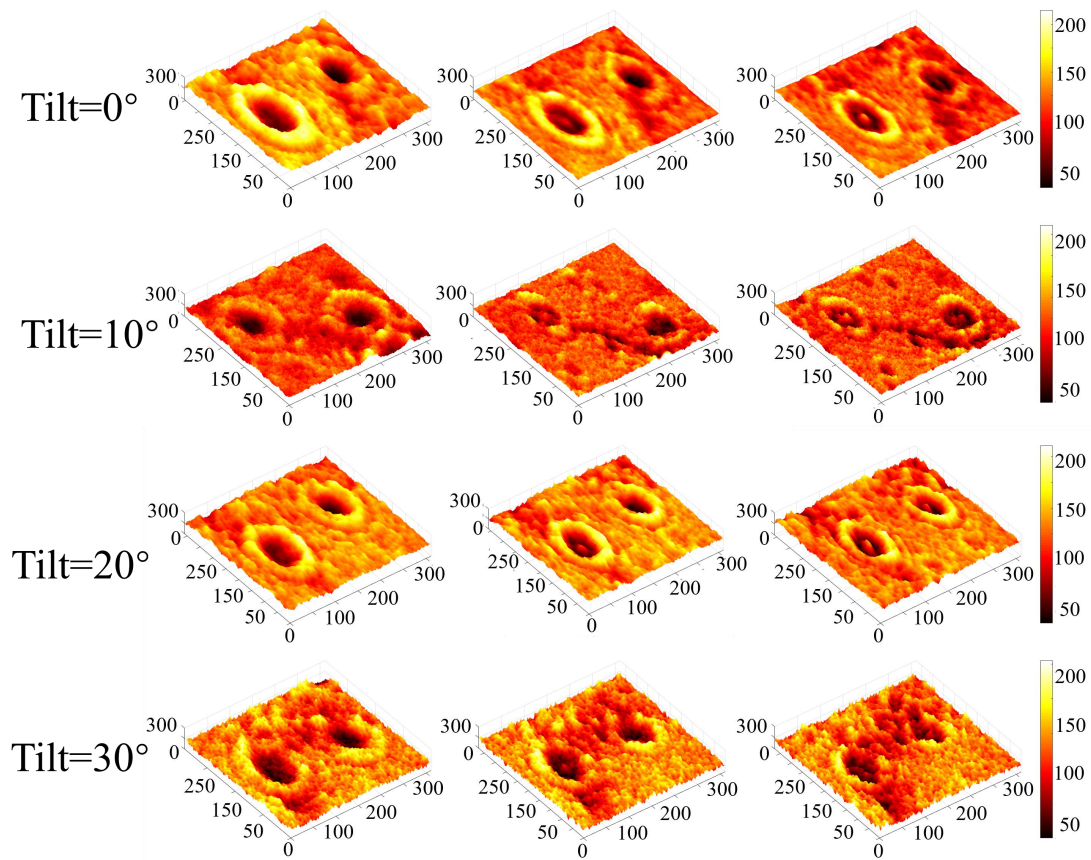

Fig.S6 Multi-skyrmion changes asynchronously at different angles.